\documentclass[12pt]{article}
\usepackage{osa2}
\usepackage{overcite}  

\begin{document}
\title{Optical Communication Noise Rejection Using Correlated Photons}

\author{Deborah Jackson, George Hockney, Jonathan Dowling}
\address{Quantum Computing Technologies Group\\ Jet Propulsion Laboratory 
\\ California 
Institute of Technology\\ 4800 Oak Grove Drive, Pasadena, California 
91109-8099}
\email{Deborah.Jackson@jpl.nasa.gov}

\begin{abstract}
This paper describes a completely new way to perform noise rejection using 
a two-photon sensitive detector and taking advantage of the properties of 
correlated photons to improve an optical 
communications link in the presence of uncorrelated noise.  In particular, 
a detailed analysis is made of the case where a classical link would be 
saturated 
by an intense background, such as when a satellite is in front of the sun, 
and 
identifies a regime where the quantum correlating system has superior 
performance.  
\end{abstract}

\section{Introduction}
The last few years have brought rapid advances in the understanding and 
application of quantum entangled photons.  Exploitation of quantum 
correlations is important in various applications, including teleportation 
\cite{Bennett95}, quantum lithography \cite{Kok02},  clock 
synchronization 
\cite{Jozsa00}, and optical communications \cite{Hong85}. To date, 
most of this work has focused on characterizing quantum correlation 
effects at low intensities due to the lack of bright sources of 
correlated photon pairs.  However, Lamas-Linares et \ al.'s 
\cite{Lamas-Linares01} recent 
report of the laser-like production of polarization entangled photons now 
opens up the possibility that one can take advantage of correlated effects 
for communications.   In addition, the demonstration of 
efficient parametric down conversion using a diode laser and a pair of 
solid state parametric conversion chips \cite{Sanaka01}, demonstrates that 
parametric 
generation technology is rapidly approaching the fabrication thresholds 
where economies of scale can be achieved.   Earlier attempts, by Mandel \cite{Hong85}, to apply 
correlated photons to improving the signal to noise ratio in a 
communications link  depends on coincidence counters, which, because of 
the detector dead time, are limited to low rates.  Our approach depends on 
the development of a special detector that is intrinsically sensitive to 
two-photons \cite{Kim99}.  This single detector replaces the coincidence counters, 
gating electronics, 
amplifiers, and the computer interface employed in that earlier effort.  Consequently, it 
is important to do a link analysis to determine whether there
is an advantage to using correlated photons and two photon 
detectors for conveying information in an optical communications channel.  

The answer is yes, but only for the special situation where the signal 
strength is swamped by in-band background noise.  We refer specifically to 
situations where free space high bandwidth of optical communications are 
desired, but in-band solar background radiation hampers the signal to 
noise.  The other situation of interest would be in fiber networks 
operating under conditions of a large background of incoherent in-band 
scattering, which adversely impacts recovery of the signal at the 
receiver.  For shot noise limited telecom links  
there is generally no advantage to using correlated photons as opposed to 
classical photons for the carrier.  Our 
approach is predicated on the development of a two photon sensitive 
photodetector which eliminates the need for separate coincidence 
electronics.   Consequently, one can define the required figure of merit 
that must be met for a performance advantage to be realzied of the 
corrrelated 2-photon link over a classical uncorrelated photon link in 
terms of the two photon and single photon quantum efficiencies of the 
detectors.  Besides introducing a new method for performing the coincidence 
measurement, this paper defines the conditions under which one can expect 
to the two-photon correlated detection to realize and advantage when 
compared to 
classical uncorrelated signal recovery.  Finally, it briefly 
examines the effect of diffraction on the 2-photon collection efficiency 
in the free space application.    

\section{Near-Field Optical Communications Links}
The near field communication link is defined as an architecture for which the full
 output power of the transmitter is subtended and collected by the 
detector.   
The schematics in Figures 1(a) and 1(b) depict the key differences between the 
classical and the correlated-photon near field telecom links.  In this 
analysis, a classical link, using photons of frequency, $\omega_{1}$, is 
compared with a correlated-photon link transmitting photon pairs of 
approximate frequency, $\omega_{1}$, but actual frequency $\omega_{a} + 
\omega_{b}$ = $2 \omega_{1}$ = $\omega_{2}$.   The power 
received at the detector is given by the following expression

\begin{equation}
P_{r}(\omega_{i},t)=\mu P_{t}(\omega_{i},t)L(\omega_{i}).
\end{equation}

\noindent where $P_{t}(\omega_{i},t)$ is the modulated output power of the
communications laser at carrier frequency $\omega_{i}$, $\mu$ represents 
conversion losses suffered after emission from the transmitter, and L
represents any losses that occur during transit through the communications
channel.  This expression assumes that the collimating optics at the 
transmitter and the collecting optics at the receiver are sufficiently 
large 
that the footprint of the transmitted beam is completely subtended by the 
receiver aperture. This would be a reasonable assumption at optical 
frequencies over free space transmission ranges up to 1000 km.  To 
facilitate comparison of the performance of both links, we assume that the 
initial laser output power of the classical link is identical to the 
laser output power of the correlated photon link, 	

\begin{equation}
P_{t}(\omega_{1},t)=P_{t}(\omega_{2},t).
\end{equation}

The link designs differ by the conversion loss factor, $\mu$, and the carrier
frequency used to transmit the signal information through the channel.  In
the classical telecom link, $\mu = 1$ and the carrier frequency is 
$\omega_{1}$.  Thus the signal current generated at the receiver for the 
classical link is given by:

\begin{equation}
I_{0}=\eta_{det}P_{r}(\omega_{1},t)=\eta_{det}P_{t}(\omega_{1},t)L(\omega_{1}),
\end{equation}

\noindent where $\eta_{det}$ is the receiver efficiency for converting input photons 
to 
carriers in the signal current.  

In contrast, the value of $\mu$ in the 
correlated photon link is determined by the method used to produce the 
correlation.  Furthermore, we assume that the modulated output of the 
laser transmitter is passed through a nonlinear parametric down conversion 
crystal and correlated 
photons are generated with efficiency, $\mu = \eta_{PDC}$.  
The following additional properties apply.
\begin{itemize}
\item  Both photons were created at the same point in time.
\item  Both photons were created at the same spatial point.
\item  Energy is conserved $\rightarrow  \omega_{2} = \omega_{a} + \omega_{b}$. 
\item  Momentum is conserved $\rightarrow  \vec{k_2} = \vec{k_a} + \vec{k_b}$. 
\end{itemize}
\noindent where one assumes that the daughter photons from the down conversion 
process are degenerate with $\omega_{a} + \omega_{b} = 2 \omega_{1} = 
\omega_{2}$.  After down conversion, 
the output is modulated with the signal waveform before being sent through 
the transmission medium for collection at the receiver.  Therefore in the 
correlated link,  
the signal current generated at the receiver is proportional to the signal 
power and is given by

\begin{equation}
I^{\prime}_{0} = \eta_{2-ph}(2\omega_{1})[P^{\prime}_{t}(\omega_{a},t) L(\omega_{a},t) 
+ 
P^{\prime}_{t}(\omega_{b},t) L(\omega_{b})] = \eta_{2-ph}(2\omega_{1}) \eta_{PDC} 
P_{t}(\omega_{2},t) L(\omega_{1}).
\end{equation}

Even though we do not know exactly when or where any pair of twin photons are born 
within the down 
conversion crystal, the fact that they 
are simultaneously created means that standard geometrical imaging optics 
can be exploited in a straightforward matter to reunite these photon pairs 
in coincidence at the receiver.  The figure of merit for comparing the 
ultimate link performance is given by the signal to noise ratio of each approach;

\begin{equation}
SNR_{classical} = \frac{I_{0}} {\sqrt{\sum_{i} \sigma^{2}_{i}}},
\end{equation}

\begin{equation}
SNR_{correlated} = \frac{I^{\prime}_{0}} {\sqrt{\sum_{i} \sigma^{2}_{i}}},
\end{equation}

where

\begin{equation}
\sum_{i} \sigma^{2}_{i} = \sigma^{2}_{thermal} + \sigma^{2}_{shot} + 
\sigma^{2}_{laserRIN} +\sigma^{2}_{background}
\end{equation}

\noindent is the sum of the variances of all noise contributions, thermal noise, 
shot noise, relative intensity noise due to the laser, and in-band 
background noise arriving with the signal at the detector.  These factors 
are defined for each link in Table 1 below.  The expression for the 
correlated SNR is optimized when the source has been imaged onto the 
detector so that the correlated pairs, that are collected by the imaging 
lens, arrive at the same time and overlap at the same point of the 
detector.  This means that correlated telecom links are practical for 
short range links where the foot print of the transmitterÕs output beam 
are fully subtended by the receiver collection aperture.

Here $k$ is BoltzmanÕs constant, $B$ is the circuit bandwidth, $T$ is the 
ambient absolute 
temperature, $R_{i}$ is the detector input impedance, $e$ is the electric 
charge, 
$F_{RIN}$ is the relative intensity noise factor, and $P_{r-B}$ is the 
power level of 
the background noise at the receiver. To compare the relative efficiencies 
of 
classical signal recovery with the signal recovery process of correlated 
photons, 
we assume similar links are established for the classical photon signal 
recovery and 
correlated photon 
signal recovery.  $P_{t}$, $A_{t}$, $R_{i}$, $L$, $A_{r}$, and 
$\omega_1$ are chosen to be identical for both 
links, where $A_{t}$ is the area of he transmiter aperture and $A_{r}$ is 
the area of the receiver collection 
aperture.  They differ in the choice of down conversion efficiencies, 
detector 
efficiencies, and the background noise contribution to the signal to noise 
ratio (SNR). 

The differences between these two types of links are summarized a 
follows.  The correlated photon transmitter design can never be as efficient as the 
classical 
transmitter due to the parametric down conversion factor, $\eta_{PDC}$ $< 1$.  
But an advantage can be obtained if one 
employs a detector that is specifically designed to have  a poor single 
photon 
detection efficiency, and a two-photon detection efficiency that is 
significantly 
larger such that $\eta_{2-ph}(2\omega_{1})$ $>>$ $\eta_{1-ph}(\omega_{1})$.  To 
illustrate this point, we 
will compare the correlated photon 
signal to noise to the classical signal to noise.  The ideal 
communications link is often 
designed so that the shot noise is dominant.  Which yields,

\begin{equation}
\frac{SNR_{correlated}}{SNR_{classical}}=\sqrt{\frac{\eta_{2-ph}(2\omega_{1})\eta_{PDC}L(\omega_{1})}{\eta_{det}}}<1.
\end{equation}

Clearly the ratio in the expression above will always be less than 1, 
largely 
because the down conversion efficiency will never be 100$\%$.  Alternatively, 
if one 
examines the case where the background noise exceeds the signal current, 
$I_{B}$ $>$ $I_{0}$,  the ratios change to

\begin{equation}
\frac{SNR_{correlated}}{SNR_{classical}}=\frac{\eta_{2-ph}(2\omega_{1})\eta_{PDC} 
L(\omega_{1})}{\sqrt{\eta_{det}\eta_{1-ph}(\omega_{1})}}>1. \label{eq:success}
\end{equation}

\noindent Here we assume that  
$\eta_{accident}(2\omega_{1})P_{r-B}$ shown in Table 1 is negligible 
compared to $\eta_{1-ph}(\omega_{1})$.  Equation~\ref{eq:success} says that the 
correlated photon link has a performance advantage as long as 
$\eta_{2-ph}(2\omega_{1})$ $>>$ $\eta_{1-ph}(\omega_{1})$, in accordance 
with this relation.   

\section{Conclusions}

Certain communications channels need to operate in the presence of severe 
background noise, such as when the Sun is positioned near the line of 
sight (LOS) between the transmitter and the receiver.  In such situations 
noise-immune coding techniques 
are of limited help due to saturation of the detector. Narrow-band filters 
can limit the background but with a sufficiently intense source the detectors can 
still 
be saturated by noise.  In this situation two-photon correlated detection 
can avoid the noise in an entirely different way, where a two-photon 
detector 
does not see it, or only sees ``accidental" coincidences which are small for 
incoherent sources like the Sun.  If this is the case a quantum-correlated 
communication channel can out-perform other techniques because it 
eliminates 
the background before a detector signal is generated.

In telecom links for which the footprint of the of the beam is larger than 
the collection aperture, quantum-correlated telecom links are essentially 
non competitive against classical communications links because the 
correlation 
cross section of the product,  $P_{t}(\omega_{a},t)P_{t}(\omega_{b},t)$, 
falls off at a rate proportional to $1/R^{4}$ while the classical link has 
a
$1/R^{2}$ dependence.  However, over ranges of about 1000 km realizable 
communications 
links using 1 to 2-meter mirror optics permit the design of links that 
collect most of 
the emitted photons, thereby making free space links under this range 
superior if the 
conditions of Equation~\ref{eq:success} are met.  In addition, fiber optic links, 
which intrinsically 
permit one to image all the incident photons onto the receiver, may find 
an advantage 
in using correlated photons when the in-band background noise exceeds 
the signal level.  

In summary, one will observe a performance 
advantage 
for correlated photon links when the classical link is limited by an 
intense source 
of uncorrelated background noise.  The requirements for using this technique are (1) 
that an 
intrinsically 
two-photon detector be developed (this rejects noise without being 
saturated by it) 
and (2) that most of the transmitted photons be collected (which can be 
done in free 
space over distances up to about 1000km).

\section*{Acknowledgements}
The research described in this paper
was carried out at the Jet Propulsion Laboratory,
California Institute of Technology, under a contract with
the National Aeronautics and
Space Administration, and was supported by a contract
with the Office of Naval Research.


\newpage
\centerline{\scalebox{.75}{\includegraphics{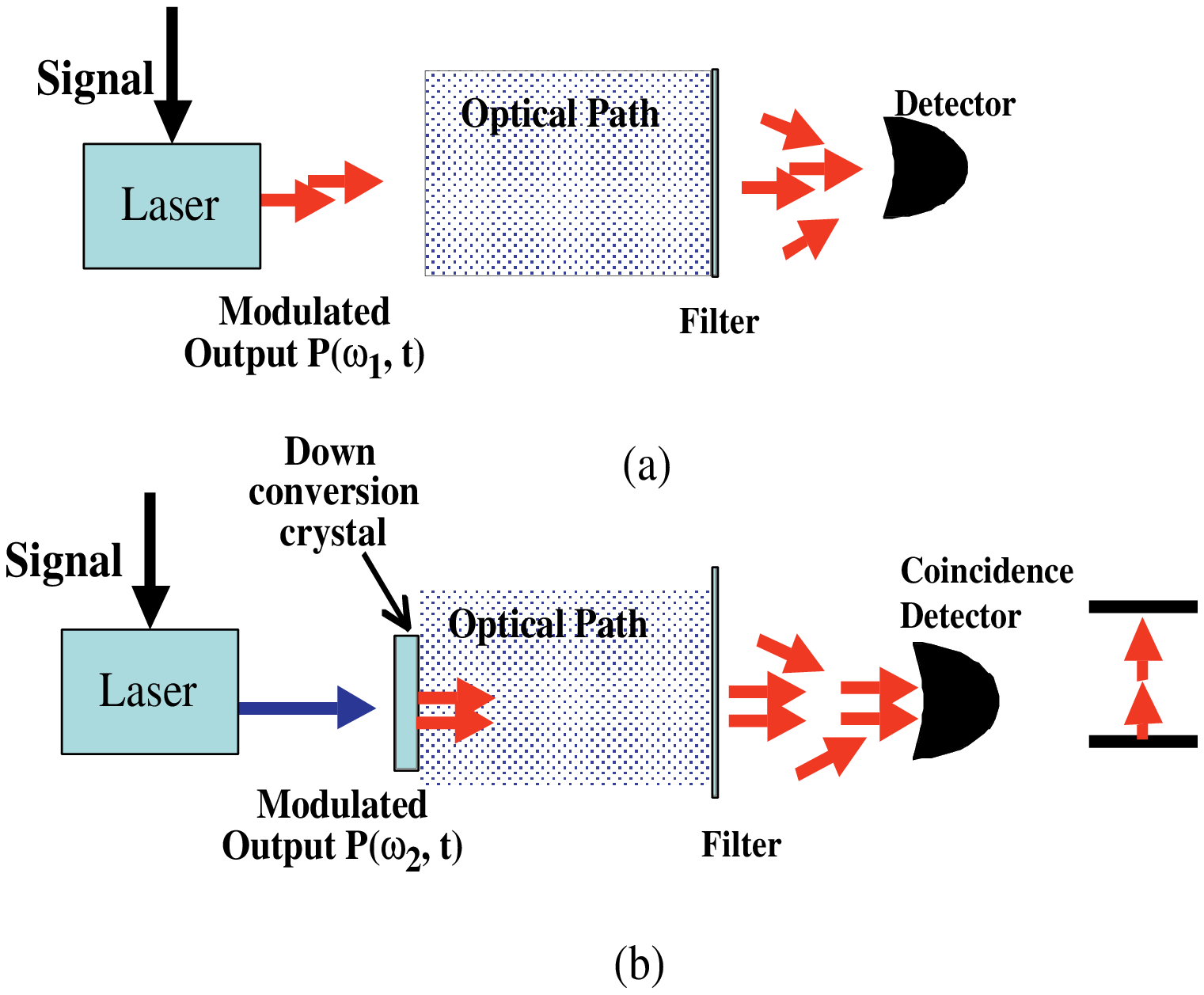}}}
\vskip10cm

Figure 1, D.J. Jackson et. al. 

\newpage
\begin{table}[h]
\vskip-.7in
\caption{Defintion of Noise Sources in Classical and Correlated Telecom 
Links}
\begin{center}
\begin{tabular}{p{1.0in}|p{1.75in}|p{3.5in}}\hline 
  Noise\ Source\ &   Classical\ Photons\ &   Correlated\ Photons\\ \hline
 $\sigma^{2}_{thermal}$\  & $\frac{8 k T B}{R_{i}}$  & $\frac{8 k T B}{R_{i}}$\ \\
 $\sigma^{2}_{shot}$\ & $4 e I_{0} B$  & $4 e I^{\prime}_{0} B$ \\
 $\sigma^{2}_{laserRIN}$\ & $2 B F_{RIN}I^{2}_{0}$  & $2 B F_{RIN} 
 (I^{\prime}_{0})^{2}$ \ \\
 $\sigma^{2}_{background}$\ & $4 e \eta_{det}P_{r-B}B$  & $4 e 
 \eta_{1-ph}(\omega_{1})P_{r-B}B$$^{\dag}$ + 
 $4e\eta_{accidental}(2\omega_{1})P_{r-B}^{2}B$$^{\ddag}$ \ \\

\hline
\end{tabular}
\end{center}
\end{table}
$\dag$ $\eta_{1-ph}(2\omega_{1})$ is the single photon detection efficiency.

$\ddag$ $\eta_{accidental}(2\omega_{1})$ is the two photon absorption 
efficiency for statistically random detection events.

\end{document}